# Viral population dynamics at the cellular level, considering the replication cycle

Seong Jun Park[a]

[a] Department of Physics and Astronomy and Center for Theoretical Physics, Seoul National University, Seoul 08826, Korea

## Abstract

Viruses are microscopic infectious agents that require a host cell for replication. Viral replication occurs in several stages, and the completion time for each stage varies due to differences in the cellular environment. Thus, the time to complete each stage in viral replication is a random variable. However, no analytic expression exists for the viral population at the cellular level when the completion time for each process constituting viral replication is a random variable. This paper presents a simplified model of viral replication, treating each stage as a renewal process with independently and identically distributed completion times. Using the proposed model, we derive an analytical formula for viral populations at the cellular level, based on viewing viral replication as a birth-death process. The mean viral count is expressed via probability density functions representing the completion time for each step in the replication process. This work validates the results with stochastic simulations. This study provides a new quantitative framework for understanding viral infection dynamics.

Email addresses: tcpsj123@cau.ac.kr (Seong Jun Park)

## 1. Introduction

For all recorded history, humans have been affected by viruses. Measles, smallpox, polio, and influenza viruses have changed the course of human history. Measles and smallpox have killed hundreds of thousands of Native Americans. Polio has killed and disabled people, including American President Franklin Delano Roosevelt. The 1918 influenza epidemic killed more people than World War I. Recently, severe acute respiratory syndrome coronavirus 2 caused coronavirus disease 2019, which has been a global problem since 2020.

According to investigations, approximately 230 virus species infect humans, causing over 200 diseases [1–4]. Further, illnesses caused by viruses account for about 10 million mortalities annually [4, 5]. The number of virus species has consistently increased with the discovery of new viruses through technological advancements [6–8]. Modern individuals are at greater risk of virus infections due to high population density, deforestation, and transportation systems. Viruses infect the target host cells to replicate their genetic material. Therefore, quantitatively understanding viral replication at the cellular level is critical.

Many studies have explored the structure of viruses and the steps of viral multiplication in cells. Viruses are classified based on several factors, including nucleic acid type (DNA or RNA), single- or double-stranded classification, and replication strategy [9–11]. Nevertheless, viruses follow the same basic and common principles when they infect a host cell. Viral replication involves several steps: attachment, entry, uncoating, replication, assembly, and release.

Over the past few decades, numerous studies have been conducted on virus infection kinetics in and between cells [12–15]. Viral replication occurs in multiple stages, but very few attempts have been made to study the number of virions in a way that reflects the well-known life cycle at the cellular level [16–18]. The completion time for each stage of viral replication fluctuates with the cell environment, encompassing the concentration of RNA polymerase or ribosomes, pH, temperature, nutritional status, and life cycle [19–23]. The completion time required for each stage of viral replication is reasonably a random variable due to fluctuation or uncertainty. The analytic expression for counting virus populations remains unresolved when the duration of each replication stage is a random variable.

While classical branching process models [24, 25] treat reproduction as an instantaneous event defined only by an offspring distribution, they ignore the temporal structure of replication. Even age-dependent extensions such as the Bellman-Harris process [26]model reproduction occurring after a single random lifetime, without resolving intermediate steps. In contrast, our framework represents the viral replication cycle as a sequence of stochastic stages with arbitrary time distributions, leading to inherently non-Markovian dynamics. This stage-resolved approach captures variability from intracellular processes and reveals how individual stages shape population-level behavior—capabilities not provided by conventional branching process models.

This paper aims to present an analytic formula for viral counting statistics at the cellular level. This work simplifies viral replication to two steps (Figure 1). Based on the replication model, we derive the analytic expression for the mean viral population using probability density functions of the completion time for each step of the model and the virus lifetime. The correctness of the analytic result is confirmed with a computational simulation. To our knowledge, this study is a novel attempt to connect the mean virus number with the completion time for each viral replication step in cells.

The paper is organized as follows. Section 2 describes the simplified viral replication model and presents the analytic expression of the mean viral population at the inter- and intracellular levels. Furthermore, we calculate the mean virus number using the analytic result, provided that the time elapsed during each process of the simple model for viral replication follows an exponential or multiterm hypo-exponential distribution. Finally, Section 3 contains the summary of this study and the research outlook.

## 2. Viral replication at the cellular level

Virus replication is the formation of a virion during the infection process in the target host cells. This formation typically occurs in seven stages [27, 28]. First, attachment occurs when viruses attach to the cell membrane of the host cell. Second, entry (penetration) is when the virus enters the host cell. Third, in uncoating, the virion protein coat (capsid) is removed, releasing its genetic material. Fourth, in replication, viruses use building blocks from the host cell (e.g., nucleotides and amino acids) to make viral proteins. Fifth, in assembly, the newly manufactured viral proteins and genomes are gathered and put together. Sixth, maturation involves capsid modifications. Finally, seventh, in release, the newly assembled and mature viruses leave the host cell.

We simplify these seven steps to two steps. First, we integrate the attachment, entry, and uncoating steps into the first step of the simplified viral replication model, called "preparation." Second, the second step of the simplified viral replication model integrates replication, assembly, maturation, and release, defined as "creation." Figure 1 presents the simplified viral replication model.

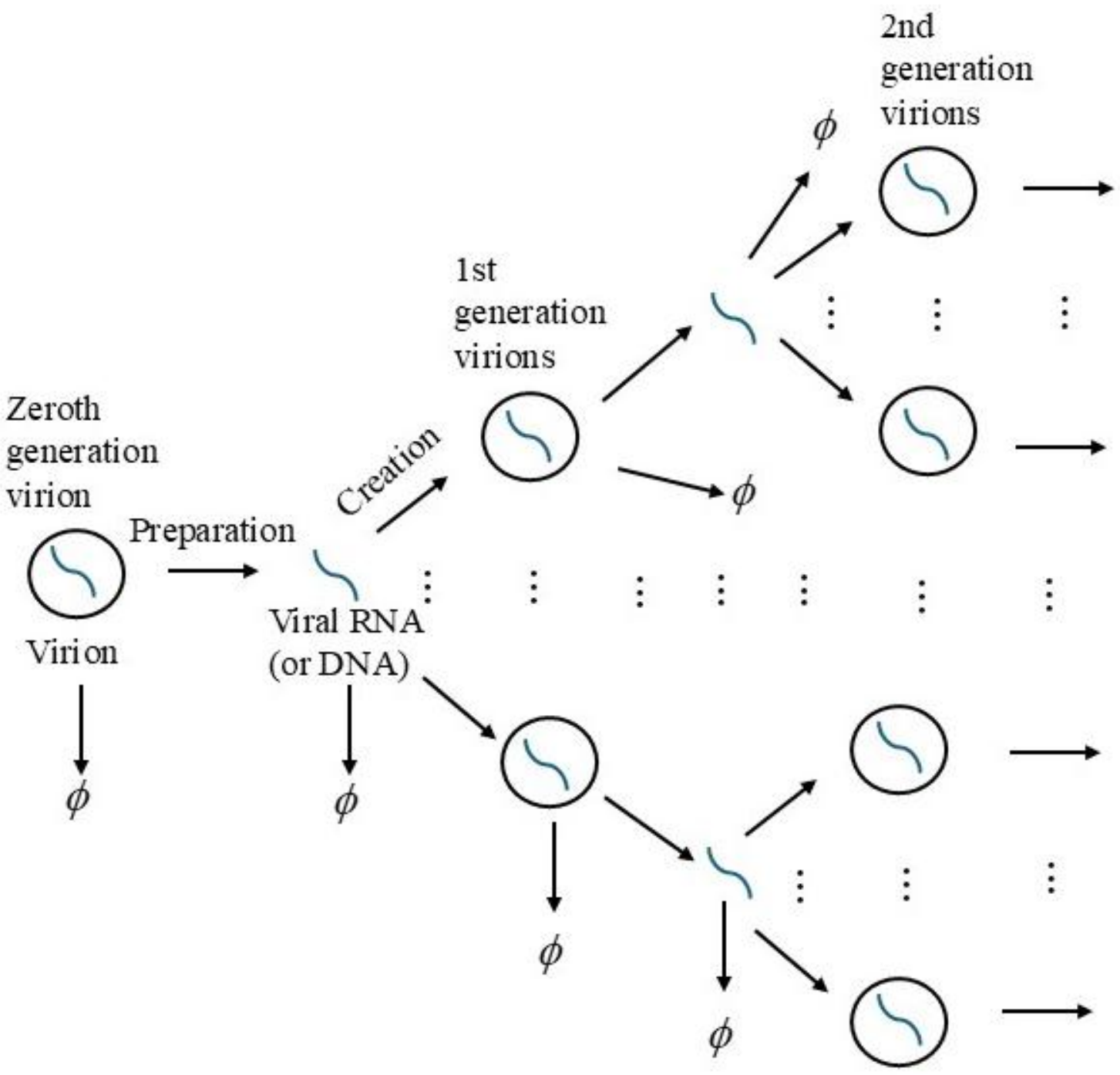

**Figure 1 Simplified virus multiplication cycle.** Viruses replicate via preparation and creation. Preparation includes the attachment of a virion to the host cell membrane, entering the host cell, and uncoating the viral genetic material (RNA or DNA). Creation includes replicating viral components, assembling into new virions, and releasing new virions to infect other cells. Preparation and creation can occur if the lifetimes of virions and viral genetic material are longer than the time needed to complete the preparation and creation steps, respectively. The $n$th generation virions are produced after $n$ rounds of replication.

Based on the simplified viral replication, we present the analytic expression of the virus number at the cellular level. Four measures of time related to viral replication are defined as follows: the time for a virion to complete the first step, preparation, denoted by $t_u$; the time at which the uncoated virus completes the second step, creation, denoted by $t_c$; the lifetime of the viral genetic material (RNA or DNA) uncoated from the preparation, denoted by $t_x$; and the lifetime of the virus, denoted by $t_d$. Because these measures of time are random variables due to their fluctuation in a complex cellular environment, we define $\varphi_u(t_u)$, $\varphi_c(t_c)$, $\varphi_x(t_x)$, and $\varphi_d(t_d)$ as the probability density functions of the four measures of time, $t_u$, $t_c$, $t_x$, and $t_d$, respectively. The complementary cumulative distribution functions of random variables ($t_u$, $t_c$, $t_x$, and $t_d$) are given by $S_u(t)=\int_t^\infty \varphi_u(\tau)d\tau$, $S_c(t)=\int_t^\infty \varphi_c(\tau)d\tau$, $S_x(t)=\int_t^\infty \varphi_x(\tau)d\tau$, and $S_d(t)=\int_t^\infty \varphi_d(\tau)d\tau$. To derive the analytic formula of the mean virus number at the cellular level, we must determine the expressions of the following quantities: 1) the probability that a virus does not degrade and complete the preparation during time interval *t*, denoted by $S_{d,u}(t)$; 2) the probability density function of the time to complete the creation without the annihilation of the viral genetic material, denoted by $\varphi_{c,x}(t)$; and 3) the probability density function of the time to complete the preparation without the death of the virus, denoted by $\varphi_{u,d}(t)$. These three terms can be expressed as $S_{d,u}(t)=S_d(t)S_u(t)$, $\varphi_{c,x}(t)=\varphi_c(t)S_x(t)$, and $\varphi_{u,d}(t)=\varphi_u(t)S_d(t)$, respectively. Using the three terms, $S_{d,u}(t)$, $\varphi_{c,x}(t)$, and $\varphi_{u,d}(t)$, and assuming that the initial virus number is 1, we obtain the mean number of viruses at the cellular level as follows:

$$\langle n(t)\rangle = S_{d,u}(t)+\sum_{i=1}^{\infty}\int_0^t S_{d,u}(\tau)\int_0^{t-\tau}\varphi_{i,u,d}(\alpha)\psi_{i,c,x}(t-\tau-\alpha)d\alpha d\tau\,, \tag{1}$$

where $\varphi_{i,u,d}(t)$ denotes the probability density function of the time to complete preparation *i* times without virus decay. Thus, this function is given by the inverse Laplace transform of $\hat{\varphi}^i_{u,d}(s)$, $\varphi_{i,u,d}(t)=L^{-1}\left(\hat{\varphi}^i_{u,d}(s)\right)$, $\hat{f}(s)\equiv\int_0^\infty f(t)e^{-st}dt$, where $L^{-1}$ represents the inverse Laplace transform. The integrand $\psi_{i,c,x}(t)$ in Eq. (1) is defined as $\psi_{i,c,x}(t)=L^{-1}\left(\left(\sum_{j=1}^{\infty}\hat{\varphi}_{j,c,x}(s)\right)^i\right)$, and $\varphi_{j,c,x}(t)$ represents the probability density function of the time until *j* creations are completed without annihilation of the viral genetic material. Moreover, $\varphi_{j,c,x}(t) = \varphi_{j,c}(t) S_x(t)$, where $\varphi_{j,c}(t)$ denotes the probability density function of the time to create *j* virions or the *j*-fold convolution of $\varphi_c(t)$ (for a detailed derivation of Eq. (1), see the Supplementary Material).

In virus population dynamics, researchers aim to determine the behavior of the viral population after an extended period. We are interested in the steady-state mean virus number. Using $\langle n(\infty) \rangle = \lim_{s \to 0} s \langle \hat{n}(s) \rangle$ and Eq. (1), we can easily derive the expected number of viruses in a steady state:

$$\langle n \rangle \equiv \langle n(\infty) \rangle = \lim_{s \to 0} \frac{s \hat{S}_{d,u}(s)}{1 - \hat{\varphi}_{u,d}(s) \hat{\psi}_{c,x}(s)}, \tag{2}$$

where $\hat{\psi}_{c,x}(s) = \sum_{j=1}^{\infty} \hat{\varphi}_{j,c,x}(s)$. A critical feature of the analytic formulas (Eqs. (1) and (2)) is that the mean virus count at the cellular level can be expressed using arbitrary probability density functions for the four time variables: preparation time, creation time, and the lifetimes of the viral genetic material and virus, reflecting the viral replication mechanism.

This work presents the results of the calculations on Eq. (1), assuming that the four measures of time (preparation, creation, and lifetimes of the viral genetic material and virus) have an exponential distribution with mean values of $\frac{1}{k_u}$, $\frac{1}{k_c}$, $\frac{1}{k_x}$, and $\frac{1}{k_d}$, respectively. Thus, their probability density functions are $\varphi_u(t) = k_u e^{-k_u t}$, $\varphi_c(t) = k_c e^{-k_c t}$, $\varphi_x(t) = k_x e^{-k_x t}$, and $\varphi_d(t) = k_d e^{-k_d t}$, respectively. The substitution of these exponential distributions into Eq. (1) yields

$$\langle n(t) \rangle = e^{-\frac{1}{2}(k_d + k_u + k_x)t} \frac{f(k_u,k_c,k_x,k_d)\cosh\left(\frac{f(k_u,k_c,k_x,k_d)t}{2}\right) + (k_x - k_d - k_u)\sinh\left(\frac{f(k_u,k_c,k_x,k_d)t}{2}\right)}{f(k_u,k_c,k_x,k_d)}, \tag{3}$$

where $f(k_u,k_c,k_x,k_d) = \sqrt{(k_x - k_d - k_u)^2 + 4k_c k_u}$, and $\cosh y = \frac{e^y + e^{-y}}{2}$ and $\sinh y = \frac{e^y - e^{-y}}{2}$ denote the hyperbolic cosine and sine functions, respectively. From Eq. (3), the mean number of virions in the steady state converges or diverges depending on the value of $f(k_u,k_c,k_x,k_d)-(k_d+k_u+k_x)$:

$$\langle n(\infty)\rangle=\begin{cases}\infty & f(k_u,k_c,k_x,k_d)-(k_d+k_u+k_x)>0\\ \frac{1}{2}\left(1+\frac{k_x-k_d-k_u}{f(k_u,k_c,k_x,k_d)}\right) & f(k_u,k_c,k_x,k_d)-(k_d+k_u+k_x)=0\\ 0 & f(k_u,k_c,k_x,k_d)-(k_d+k_u+k_x)<0\,.\end{cases}\qquad(4)$$

The assumption that the four measures of time describing virion kinetics have exponential distributions is a manifestation of the single step and constant rate of each process (i.e., preparation, creation, death of virions, and death of viral genetic material). However, these rates are not constant because they fluctuate with time and the cellular environment (e.g., pH, temperature, and nutritional status), and each process includes multiple steps to complete [14, 19–23, 29, 30]. The rates vary when each process for the birth and death of the virions occurs. This work focuses on the preparation process in which virion production comprises multiple processes, including attachment, entry, and uncoating. Then, we consider the preparation time to be a multiterm hypo-exponential distribution, $\varphi_u(t)=\left(\prod_{\alpha=1}^{m}k_\alpha\right)\sum_{r=1}^{m}\frac{e^{-k_r t}}{\prod_{\substack{\beta=1\\ \beta\neq r}}^{m}(k_\beta-k_r)}$, where $m$ denotes the number of processes in the preparation process, $k_r$ ($r=1,2,\cdots,m$) represents the rate (distinctly positive constants) of the process involved in the preparation process, and $\varphi_u(t)$ satisfies the property of probability density function $\int_0^\infty\varphi_u(t)dt=1$. The creation time and lifetimes of the viral DNA and virions have exponential distributions, as mentioned. We write the following Laplace transform for the virion number:

$$\langle n(t)\rangle=L^{-1}\left(\frac{\sum_{r=1}^{m}\frac{c_r}{k_r(s+k_r+k_d)}}{1-\frac{k_c}{s+k_x}\prod_{r=1}^{m}\frac{c_r}{s+k_r+k_d}}\right),\qquad(5)$$

where $L^{-1}$ denotes the inverse Laplace transform, and $c_r = \frac{\prod_{\alpha=1}^{m} k_\alpha}{\prod_{\substack{\beta=1 \\ \beta \neq r}}^{m} (k_\beta - k_r)}$. The mean number of virions in Eq. (5) is a multiexponential function that is a linear combination of the exponential functions. Figures 2 and 3 reveal that the analytic results (Eqs. (3)–(5)) are consistent with the stochastic simulations. The analytic expression of the mean virion number at the cellular level in Eq. (1) reveals the relationship between the four measures of time and the mean virion number.

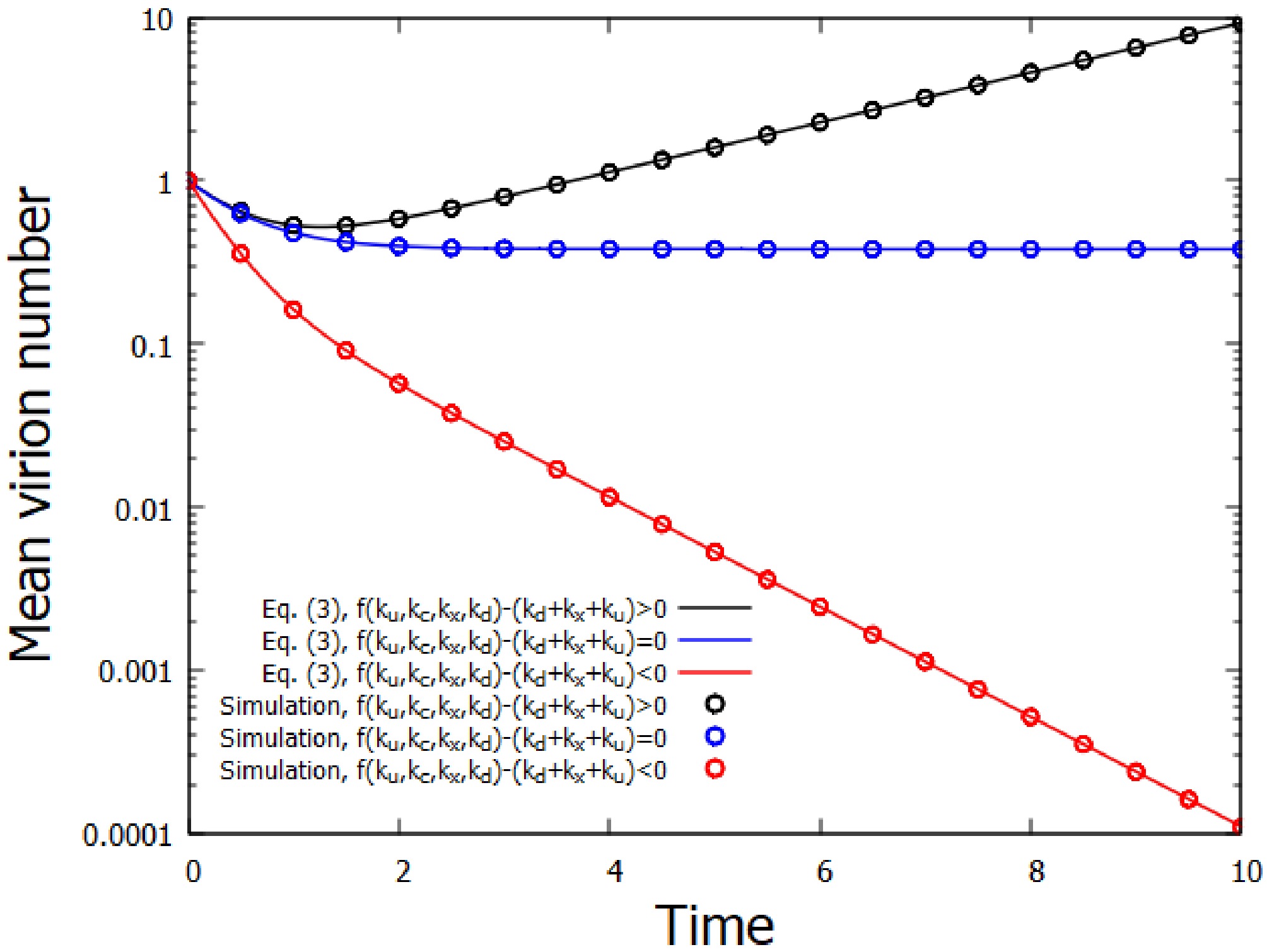

**Figure 2 Mean virion number with exponentially distributed preparation time, creation time, and lifetimes of the viral genetic material and virion.** The lines and circles respectively indicate the theoretical results for Eq. (3) and the stochastic simulation, where $k_u$ denotes the preparation rate, $k_c$ indicates the creation rate, $k_x$ represents the degradation rate of the viral genetic material, and $k_d$ signifies the degradation rate of a virion. The black line and circles represents $f(k_u, k_c, k_x, k_d) - (k_d + k_x + k_u) > 0$, where $f(k_u, k_c, k_x, k_d) = \sqrt{(k_x - k_d - k_u)^2 + 4k_c k_u}$. Over time, the mean virion number becomes greater when $f(k_u, k_c, k_x, k_d) - (k_d + k_x + k_u) > 0$ ($k_u = 0.8$,

$k_c = 1.0$, $k_x = 1.3$, and $k_d = 1.5$ on the black line with circles). The blue line and circles correspond to the mean virion number when $f(k_u, k_c, k_x, k_d) - (k_d + k_x + k_u) = 0$ ($k_u = 0.8$, $k_c = 1.0$, $k_x = 0.7$, and $k_d = 0.343$ on the blue line and circles) and reveals that the mean virion number reaches the saturation level (0.380), consistent with Eq. (4). The mean virion number when $f(k_u, k_c, k_x, k_d) - (k_d + k_x + k_u) < 0$ is marked by a red line and circles ($k_u = 0.8$, $k_c = 1.0$, $k_x = 0.2$ and $k_d = 0.3$). According to the red line and circles, the mean virion number converges to zero as time approaches infinity.

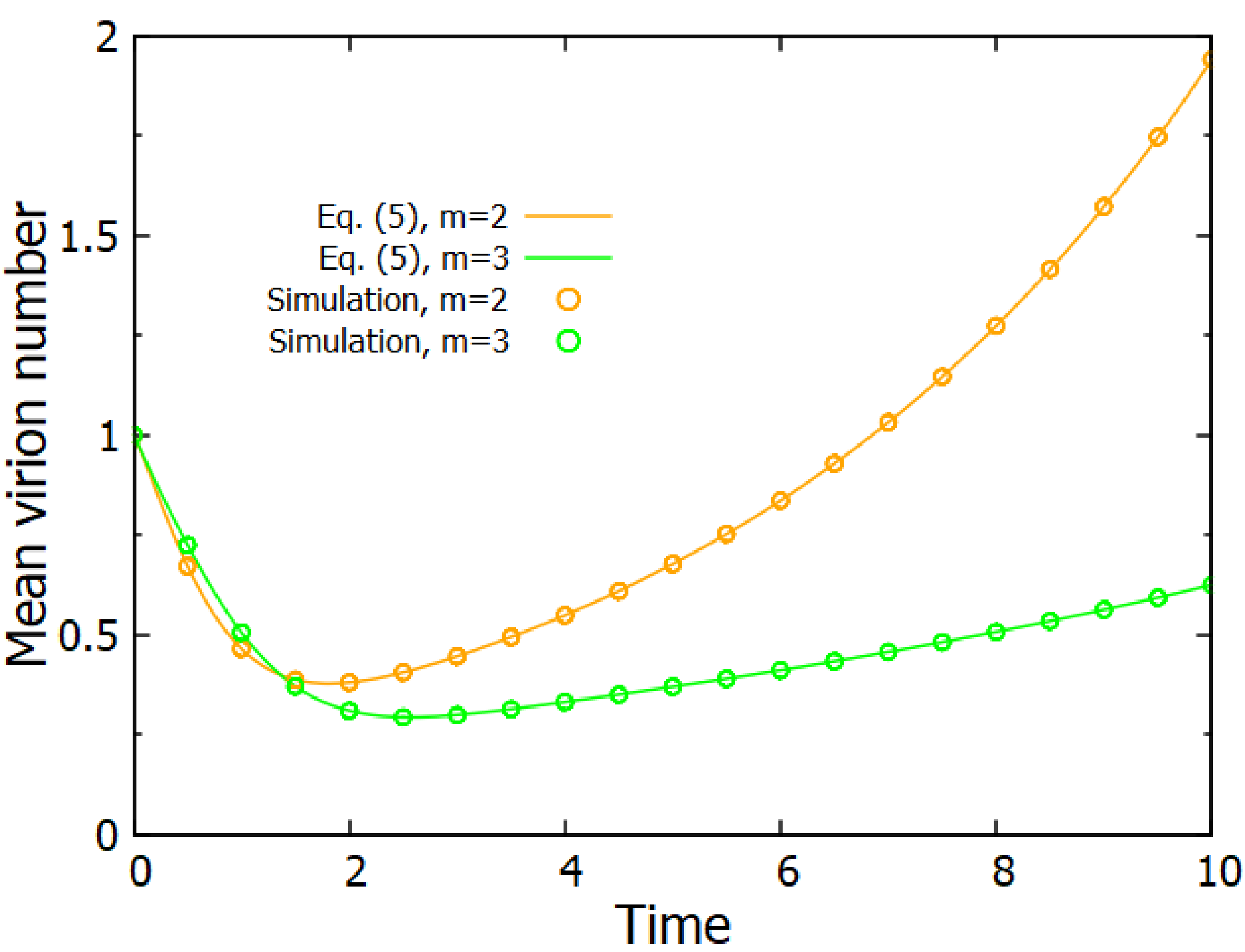

**Figure 3 Mean virion number with the preparation process comprising multiple steps.** The lines and circles are obtained from Eq. (5) and the computational simulations, respectively. The orange line and circles correspond to the mean virion number when the number of processes ($m$) in the preparation process is 2. The parameter values are $k_1 = 2.0$, $k_2 = 0.75$, $k_c = 1.2$, $k_x = 0.2$, and $k_d = 0.6$ for the orange line with circles. The mean virion number from $m = 3$ is marked by the green line and circles. The parameter values for the green line with circles are identical to those of the orange line and circles, except for the inclusion of an additional parameter, $k_3 = 1.4$, in the green set.

## 3. Summary and outlook

This paper proposes a simplified model for viral replication and an analytic expression of the mean virion number at the cellular level. The simplified model appears in Fig. 1, which indicates that the following four measures of time are necessary to describe viral replication: preparation time for a virion to enter a host cell and uncover its genetic material, creation time in which viral DNA produces virions using gene expression machinery from the host cell, the lifetime of the viral DNA, and the lifetime of the virions. These measures of time can be considered random variables due to their variability originating from the cell environment (e.g., nutritional condition and life cycle) [19–23]. This work presents the analytic expression, Eq. (1), of the mean virions, and Fig. 1 presents the schema of these dynamics.

The primary contribution of this study is to obtain analytic results, Eqs. (1) and (2), of the mean virion number based on the arbitrary probability density functions of the four measures of time. We compute the mean number of virions, Eqs. (3) and (4), given that the four measures of time have exponential distributions. The preparation process in the simplified viral replication model encompasses many stages, including attachment, entry, and uncoating. This work reflects the multistep preparation process and calculates the mean number of virions, Eq. (5), if the preparation time has a multiterm hypo-exponential distribution and the rest of the measures of time have an exponential distribution. Figures 2 and 3 indicate that the analytic results are correct, aligning with the stochastic simulations.

The analytic result, Eq. (1), holds in all viruses because the simplified replication model of a virus (Fig. 1) includes the essential processes in which viruses enter a host cell and employ the cell to reproduce. We can quantitatively understand the relationship between the elapsed time during each viral replication stage and the mean number of virions in the transient and steady states, Eqs. (1) and (2). Such a quantitative understanding provides the theoretical basis for how much time it takes to complete the viral replication process. For example, scientists can develop effective antiviral drugs to regulate the process that requires the most time in viral replication.

We can extend the framework to derive Eq. (1) for refined viral replication models. Viruses are primarily classified using the Baltimore classification system [9, 28], which categorizes them into seven groups based on the type of genetic material (DNA or RNA) and the method of producing messenger RNA. The detailed replication mechanisms of viruses differ depending on the seven classification groups.

Next, future work should consider viruses that attack bacteria, called bacteriophages. Two replication cycles reproduce bacteriophages, the lytic and lysogenic cycles [31, 32]. These cycles have their own mechanisms to produce bacteriophages. In future work, we aim to establish more sophisticated and specific replication models for viruses categorized by the Baltimore classification system and bacteriophages. We also regard the time to complete each process of these new viral replication models as a random variable. Moreover, we aim to derive analytic expressions of the mean number of virions that follow the replication mechanism. The analytic results of the mean virion numbers are helpful for understanding viral replication at the cellular level in a detailed and quantitative manner.

**Declaration of Competing Interest**

The authors declare that they have no known competing financial interests or personal relationships that could have appeared to influence the work reported in this paper.

**Data Availability**

No data were used for the research described in the article.

**Acknowledgment**

This work was supported by a National Research Foundation (NRF) grant funded by the Korean government (Ministry of Science and ICT) [Grant No. RS2021NR061970 (SJP)].

**CRediT authorship contribution statement**

Seong Jun Park: Conceptualization, Methodology, Study design, Software, Validation, Formal analysis, Investigation, Resources, Data curation, Writing – original draft, Writing – review and editing, Visualization, Supervision, Project administration, Funding acquisition

## Supplementary Material

## Viral population dynamics at the cellular level, considering the replication cycle

### Derivation of Eq. (1)

We examine the number of virions at the inter- and intracellular levels. Figure 1 describes the virion dynamics. This work assumes that an initial virion multiplies by infecting a host cell and uses the cell machinery to copy its genetic material. To reproduce, viruses disassemble, releasing their nucleic acids and proteins. Then, new virions are formed and released from the host cell. Next, the virions try to infect other cells. Based on the virus life cycle, this work defines four measures of time: the virus lifetime ($\tau_d$), the time for a virion to enter a host cell and remove its capsid that protects the genetic material ($\tau_u$) of the virus, the time for the viral genetic material to produce a new virion ($\tau_c$), and the lifetime of the viral genetic material (DNA or RNA) ($\tau_x$).

When the initial virion infects the host cell, the new virions produced in the cell are the first generation. First-generation virions produce second-generation virions, either in the same cell or in other cells they infect. Theoretically, the process can be continued.

Next, we examine the cases contributing to the number of virions. If the initial virion survives without death and uncoating, this case counts in the number of virions. Then, we consider the virions of the first generation. The initial virion must take its capsid off, which is called uncoating. The uncoated virus replicates and produces new virions. The first-generation virions that do not undergo uncoating and death can be counted in the number of virions. The *i*th generation ($i = 1, 2, 3, \cdots$) can contribute to the virion population unless they release the capsid or decay. Thus, the number of virions at time *t* is as follows:

$$
\begin{aligned}
n(t) = {} & \theta(\tau_0^d - t)\theta(\tau_0^u - t) + \theta(\tau_0^d - \tau_0^u)\sum_{j_1=1}^{\infty} \theta(t - t_{1,j_1}^c)\theta(t_{1,j_1}^d - t)\theta(t_{1,j_1}^u - t)\theta\left(\tau_0^x - \sum_{l=1}^{j_1} \tau_{1,l}^c\right) \\
& + \theta(\tau_0^d - \tau_0^u)\sum_{i=2}^{\infty}\sum_{j_1=1}^{\infty}\sum_{j_2=1}^{\infty}\cdots\sum_{j_i=1}^{\infty}\left\{\begin{aligned} & \left(\prod_{m=1}^{i-1}\theta(\tau_{m,j_m}^d - \tau_{m,j_m}^u)\right)\theta(t - t_{i,j_i}^c)\theta(t_{i,j_i}^u - t)\theta(t_{i,j_i}^d - t) \\ & \left(\prod_{q=0}^{i-2}\theta\left(\tau_{q,j_q}^x - \sum_{r=1}^{b_{q+1}}\tau_{q+1,y_{r,q+1}}^c\right)\right)\theta\left(\tau_{\mathrm{i}-1,j_{i-1}}^x - \sum_{l=1}^{g}\tau_{i,z_l}^c\right) \end{aligned}\right\},
\end{aligned} \tag{A-1}
$$

where $\theta(x)$ denotes the Heaviside step function, $\tau_0^d$ and $\tau_0^u$ represent the time until death and the uncoating of the initial virion, respectively. In addition, $\tau_0^x$ indicates the lifetime of the uncoated genetic material of the initial virion, $\tau_{i,j}^x$ denotes the lifetime of the uncoated genetic material of the *j*th virion of the *i*th generation, $\tau_{i,j}^c$ signifies the time for the uncoated viral genetic material to produce *j* virions of the *i*th generation, $\tau_{i,j}^d$ represents the lifetime of the *j*th virion in the *i*th generation, $\tau_{i,j}^u$ indicates the time to remove the capsid of the *j*th virion in the *i*th generation, $t_{i,j}^c$ is the time until the *j*th virion in the *i*th generation, $t_{i,j}^d$ is the time to degrade the *j*th virion in the *i*th generation, and $t_{i,j}^u$ indicates the time to eliminate the capsid of the *j*th virion in the *i*th generation. By definition, $t_{i,j}^c = \tau_0^u + \sum_{m=1}^{i-1}(\tau_{m,j_m}^u + \tau_{m,j_m}^c) + \tau_{i,j}^c$, $t_{i,j}^u = t_{i,j}^c + \tau_{i,j}^u$, and $t_{i,j}^d = t_{i,j}^c + \tau_{i,j}^d$. In the last term of Eq. (1), $\tau_{0,j_0}^x = \tau_0^x$, the upper bound of the summations $b_{q+1}$ and $g$ represents the number of virions produced from the (*j*-1)th produced virions in the (*j*-1)th generation, $y_{b_{q+1},q+1} = j_{q+1}$ and $z_g = j_i$.

To derive the mean number of virions, the next step is to determine the mean of the virion population. We must take an average of the random variables $\{\tau_0^d, \tau_0^u, \tau_0^x, \tau_{i,j}^c, \tau_{i,j}^x, \tau_{i,j}^d, \tau_{i,j}^u; i = 1,2,\cdots, j = 1,2,\cdots\}$ in Eq. (A-1). The probability density functions of $\tau_0^d$ and $\tau_{i,j}^d$ are defined by $\varphi_d(y)$ ($y \in \{\tau_0^d, \tau_{i,j}^d\}$). The probability density function of $\tau_{i,j}^c$ is defined by $\varphi_c(\tau_{i,j}^c)$. The probability density functions of $\tau_0^x$ and $\tau_{i,j}^x$ are defined by $\varphi_x(z)$ ($z \in \{\tau_0^x, \tau_{i,j}^x\}$), and the probability density functions of $\tau_0^u$ and $\tau_{i,j}^u$ are defined by $\varphi_u(b)$ ($b \in \{\tau_0^u, \tau_{i,j}^u\}$). Focusing on the first term on the right-hand side (RHS) of Eq. (A-1), we take the average over $\{\tau_0^d, \tau_0^u\}$:

$$
\begin{aligned}
\langle \theta(\tau_0^d - t)\theta(\tau_0^u - t)\rangle &= \int_0^\infty \int_0^\infty \theta(\tau_0^d - t)\theta(\tau_0^u - t)\varphi_d(\tau_0^d)\varphi_u(\tau_0^u)d\tau_0^u d\tau_0^d \\
&= \int_t^\infty \varphi_d(\tau_0^d)d\tau_0^d \int_t^\infty \varphi_u(\tau_0^u)d\tau_0^u \\
&= S_d(t)S_u(t)
\end{aligned}
\qquad \text{(A-2)}
$$

,

where $S_d(t) = \int_t^\infty \varphi_d(x)dx$ and $S_u(t) = \int_t^\infty \varphi_u(x)dx$ represent the complementary cumulative distribution functions of the lifetime of virions and the time until uncoating after virion creation, respectively. Next, for the second term on the RHS of Eq. (A-1), as $t_{1,1}^c = \tau_0^u + \tau_{1,1}^c$, $t_{1,1}^d = t_{1,1}^c + \tau_{1,1}^d$, and $t_{1,1}^u = t_{1,1}^c + \tau_{1,1}^u$, it follows that

$$
\begin{aligned}
&\langle \theta(\tau_0^d - \tau_0^u)\theta(t - t_{1,1}^c)\theta(t_{1,1}^d - t)\theta(t_{1,1}^u - t)\theta(\tau_0^x - \tau_{1,1}^c)\rangle \\
&= \langle \theta(\tau_0^d - \tau_0^u)\theta(t - \tau_0^u - \tau_{1,1}^c)\theta(\tau_0^u + \tau_{1,1}^c + \tau_{1,1}^d - t)\theta(\tau_0^u + \tau_{1,1}^c + \tau_{1,1}^u - t)\theta(\tau_0^x - \tau_{1,1}^c)\rangle \\
&= \int_0^\infty \int_0^\infty \int_0^\infty \int_0^\infty \int_0^\infty \int_0^\infty \theta(\tau_0^d - \tau_0^u)\theta(t - \tau_0^u - \tau_{1,1}^c)\theta(\tau_0^u + \tau_{1,1}^c + \tau_{1,1}^d - t)\theta(\tau_0^u + \tau_{1,1}^c + \tau_{1,1}^u - t)\theta(\tau_0^x - \tau_{1,1}^c) \\
&\qquad \varphi_d(\tau_0^d)\varphi_u(\tau_0^u)\varphi_c(\tau_{1,1}^c)\varphi_d(\tau_{1,1}^d)\varphi_u(\tau_{1,1}^u)\varphi_x(\tau_{1,1}^x)d\tau_0^d d\tau_0^u d\tau_{1,1}^c d\tau_{1,1}^d d\tau_{1,1}^u d\tau_{1,1}^x \\
&= \int_0^t \int_0^\infty \int_0^\infty \int_0^\infty \int_0^\infty \int_0^\infty \int_0^\infty \delta(\tau - \tau_0^u - \tau_{1,1}^c)\theta(\tau_0^d - \tau_0^u)\theta(\tau_0^u + \tau_{1,1}^c + \tau_{1,1}^d - t)\theta(\tau_0^u + \tau_{1,1}^c + \tau_{1,1}^u - t)\theta(\tau_0^x - \tau_{1,1}^c) \\
&\qquad \varphi_d(\tau_0^d)\varphi_u(\tau_0^u)\varphi_c(\tau_{1,1}^c)\varphi_d(\tau_{1,1}^d)\varphi_u(\tau_{1,1}^u)\varphi_x(\tau_{1,1}^x)d\tau_0^d d\tau_0^u d\tau_{1,1}^c d\tau_{1,1}^d d\tau_{1,1}^u d\tau_{1,1}^x d\tau \\
&= \int_0^t \int_0^\infty \int_0^\infty \int_0^\infty \int_0^\infty \int_0^\infty \delta(\tau - \tau_0^u - \tau_{1,1}^c)S_d(\tau_0^u)\theta(\tau_0^u + \tau_{1,1}^c + \tau_{1,1}^d - t)\theta(\tau_0^u + \tau_{1,1}^c + \tau_{1,1}^u - t)\theta(\tau_0^x - \tau_{1,1}^c) \\
&\qquad \varphi_u(\tau_0^u)\varphi_c(\tau_{1,1}^c)\varphi_d(\tau_{1,1}^d)\varphi_u(\tau_{1,1}^u)\varphi_x(\tau_{1,1}^x)d\tau_0^u d\tau_{1,1}^c d\tau_{1,1}^d d\tau_{1,1}^u d\tau_{1,1}^x d\tau \\
&= \int_0^t \int_0^\infty \int_0^\infty \int_0^\infty \int_0^\infty \int_0^y \delta(\tau - y)S_d(y - \tau_{1,1}^c)\theta(y + \tau_{1,1}^d - t)\theta(y + \tau_{1,1}^u - t)\theta(\tau_0^x - \tau_{1,1}^c) \\
&\qquad \varphi_u(y - \tau_{1,1}^c)\varphi_c(\tau_{1,1}^c)\varphi_d(\tau_{1,1}^d)\varphi_u(\tau_{1,1}^u)d\tau_{1,1}^c dy d\tau_0^d d\tau_{1,1}^u d\tau_0^x d\tau \\
&= \int_0^t \int_0^\infty \int_0^\infty \int_0^\infty \int_0^\tau S_d(\tau - \tau_{1,1}^c)\theta(\tau + \tau_{1,1}^d - t)\theta(\tau + \tau_{1,1}^u - t)\theta(\tau_0^x - \tau_{1,1}^c) \\
&\qquad \varphi_u(\tau - \tau_{1,1}^c)\varphi_c(\tau_{1,1}^c)\varphi_d(\tau_{1,1}^d)\varphi_u(\tau_{1,1}^u)d\tau_{1,1}^c d\tau_0^d d\tau_{1,1}^u d\tau_0^x \, d\tau \\
&= \int_0^t \int_0^\infty \int_0^\infty \int_0^\tau S_d(\tau - \tau_{1,1}^c)\theta\left(\tau_{1,1}^d - (t - \tau)\right)\theta\left(\tau_{1,1}^u - (t - \tau)\right)S_x(\tau_{1,1}^c) \\
&\qquad \varphi_u(\tau - \tau_{1,1}^c)\varphi_c(\tau_{1,1}^c)\varphi_d(\tau_{1,1}^d)\varphi_u(\tau_{1,1}^u)d\tau_{1,1}^c d\tau_{1,1}^d d\tau_{1,1}^u d\tau \\
&= \int_0^t \int_0^\tau S_d(\tau - \tau_{1,1}^c)S_d(t - \tau)S_u(t - \tau)S_x(\tau_{1,1}^c)\varphi_u(\tau - \tau_{1,1}^c)\varphi_c(\tau_{1,1}^c)d\tau_{1,1}^c d\tau \\
&= \int_0^t \int_0^\tau \varphi_u(z)S_d(z)\varphi_c(\tau - z)S_x(\tau - z)dz \, S_d(t - \tau)S_u(t - \tau)d\tau
\end{aligned}
\qquad \text{(A-3)}
$$

where $\delta(x)$ represents the Dirac delta function. The third equality uses $\theta(t) = \int_0^t \delta(\tau) d\tau$, and the fifth equality follows by the change of variable, $\tau_0^u + \tau_{1,1}^c = y$. The sixth equality is obtained using the identity $\int_0^\infty \delta(\tau - x) f(x) dx = f(\tau)$, and $S_x(t)$ in the seventh equality is the complementary cumulative distribution function of the lifetime of the uncoated viral genetic material ($\tau_x$), $S_x(t) = \int_t^\infty \varphi_x(\tau) d\tau$. Then, we consider $\langle \theta(\tau_0^d - \tau_0^u)\theta(t - t_{1,2}^c)\theta(t_{1,2}^d - t)\theta(t_{1,2}^u - t)\theta(\tau_0^x - \tau_{1,1}^c - \tau_{1,2}^c) \rangle$. As $t_{1,2}^c = \tau_0^u + \tau_{1,1}^c + \tau_{1,2}^c$, $t_{1,2}^d = t_{1,2}^c + \tau_{1,2}^d$, and $t_{1,2}^u = t_{1,2}^c + \tau_{1,2}^u$, we obtain

$$
\begin{aligned}
&\langle \theta(\tau_0^d - \tau_0^u)\theta(t - t_{1,2}^c)\theta(t_{1,2}^d - t)\theta(t_{1,2}^u - t)\theta(\tau_0^x - \tau_{1,1}^c - \tau_{1,2}^c) \rangle \\
&= \langle \theta(\tau_0^d - \tau_0^u)\theta(t - \tau_0^u - \tau_{1,1}^c - \tau_{1,2}^c)\theta(\tau_0^u + \tau_{1,1}^c + \tau_{1,2}^c + \tau_{1,2}^d - t)\theta(\tau_0^u + \tau_{1,1}^c + \tau_{1,2}^c + \tau_{1,2}^u - t)\theta(\tau_0^x - \tau_{1,1}^c - \tau_{1,2}^c) \rangle \\
&= \int_0^\infty \int_0^\infty \int_0^\infty \int_0^\infty \int_0^\infty \int_0^\infty \int_0^\infty \left( \begin{array}{l} \theta(\tau_0^d - \tau_0^u)\theta(t - \tau_0^u - \tau_{1,1}^c - \tau_{1,2}^c)\theta(\tau_0^u + \tau_{1,1}^c + \tau_{1,2}^c + \tau_{1,2}^d - t) \\ \theta(\tau_0^u + \tau_{1,1}^c + \tau_{1,2}^c + \tau_{1,2}^u - t)\theta(\tau_0^x - \tau_{1,1}^c - \tau_{1,2}^c)\varphi_d(\tau_0^d)\varphi_u(\tau_0^u)\varphi_c(\tau_{1,1}^c)\varphi_c(\tau_{1,2}^c)\varphi_d(\tau_{1,2}^d) \\ \varphi_u(\tau_{1,2}^u)\varphi_x(\tau_{1,1}^x) d\tau_0^d d\tau_0^u d\tau_{1,1}^c d\tau_{1,2}^c d\tau_{1,2}^d d\tau_{1,2}^u d\tau_0^x \end{array} \right) \\
&= \int_0^t \int_0^\infty \int_0^\infty \int_0^\infty \int_0^\infty \int_0^\infty \int_0^\infty \int_0^\infty \left( \begin{array}{l} \delta(\tau - \tau_0^u - \tau_{1,1}^c - \tau_{1,2}^c)\theta(\tau_0^d - \tau_0^u)\theta(\tau_0^u + \tau_{1,1}^c + \tau_{1,2}^c + \tau_{1,2}^d - t) \\ \theta(\tau_0^u + \tau_{1,1}^c + \tau_{1,2}^c + \tau_{1,2}^u - t)\theta(\tau_0^x - \tau_{1,1}^c - \tau_{1,2}^c)\varphi_d(\tau_0^d)\varphi_u(\tau_0^u)\varphi_c(\tau_{1,1}^c)\varphi_c(\tau_{1,2}^c)\varphi_d(\tau_{1,2}^d) \\ \varphi_u(\tau_{1,2}^u)\varphi_x(\tau_0^x) d\tau_0^d d\tau_0^u d\tau_{1,1}^c d\tau_{1,2}^c d\tau_{1,2}^d d\tau_{1,2}^u d\tau_0^x d\tau \end{array} \right) \\
&= \int_0^t \int_0^\infty \int_0^\infty \int_0^\infty \int_0^\infty \int_0^\infty \int_0^\infty \left( \begin{array}{l} \delta(\tau - \tau_0^u - \tau_{1,1}^c - \tau_{1,2}^c) S_d(\tau_0^u)\theta(\tau_0^u + \tau_{1,1}^c + \tau_{1,2}^c + \tau_{1,2}^d - t) \\ \theta(\tau_0^u + \tau_{1,1}^c + \tau_{1,2}^c + \tau_{1,2}^u - t)\theta(\tau_0^x - \tau_{1,1}^c - \tau_{1,2}^c)\varphi_u(\tau_0^u)\varphi_c(\tau_{1,1}^c)\varphi_c(\tau_{1,2}^c) \\ \varphi_d(\tau_{1,2}^d)\varphi_u(\tau_{1,2}^u)\varphi_x(\tau_0^x) d\tau_0^u d\tau_{1,1}^c d\tau_{1,2}^c d\tau_{1,2}^d d\tau_{1,2}^u d\tau_0^x d\tau \end{array} \right) \\
&= \int_0^t \int_0^\infty \int_0^\infty \int_0^\infty \int_0^\infty \int_0^w \int_0^y \left( \begin{array}{l} \delta(\tau - w) S_d(y - \tau_{1,1}^c)\theta(w + \tau_{1,2}^d - t)\theta(w + \tau_{1,2}^u - t)\theta\left(\tau_0^x - \tau_{1,1}^c - (w - y)\right) \\ \varphi_u(y - \tau_{1,1}^c)\varphi_c(\tau_{1,1}^c)\varphi_c(w - y)\varphi_d(\tau_{1,2}^d)\varphi_u(\tau_{1,2}^u)\varphi_x(\tau_0^x) d\tau_{1,1}^c dy dw d\tau_{1,2}^d d\tau_{1,2}^u d\tau_0^x d\tau \end{array} \right) \\
&= \int_0^t \int_0^\infty \int_0^\infty \int_0^\infty \int_0^\tau \int_0^y \left( \begin{array}{l} S_d(y - \tau_{1,1}^c)\theta(\tau + \tau_{1,2}^d - t)\theta(\tau + \tau_{1,2}^u - t)\theta\left(\tau_0^x - \tau_{1,1}^c - (\tau - y)\right)\varphi_u(y - \tau_{1,1}^c) \\ \varphi_c(\tau_{1,1}^c)\varphi_c(\tau - y)\varphi_d(\tau_{1,2}^d)\varphi_u(\tau_{1,2}^u)\varphi_x(\tau_0^x) d\tau_{1,1}^c dy d\tau_{1,2}^d d\tau_{1,2}^u d\tau_0^x d\tau \end{array} \right) \\
&= \int_0^t \int_0^\infty \int_0^\infty \int_0^\tau \int_0^y \left( \begin{array}{l} S_d(y - \tau_{1,1}^c)\theta(\tau + \tau_{1,2}^d - t)\theta(\tau + \tau_{1,2}^u - t) S_x(\tau + \tau_{1,1}^c - y)\varphi_u(y - \tau_{1,1}^c) \\ \varphi_c(\tau_{1,1}^c)\varphi_c(\tau - y)\varphi_d(\tau_{1,2}^d)\varphi_u(\tau_{1,2}^u) d\tau_{1,1}^c dy d\tau_{1,2}^d d\tau_{1,2}^u d\tau \end{array} \right) \\
&= \int_0^t \int_0^\tau \int_0^y \left( \begin{array}{l} S_d(y - \tau_{1,1}^c) S_d(t - \tau) S_u(t - \tau) S_x\left(\tau - (y - \tau_{1,1}^c)\right)\varphi_u(y - \tau_{1,1}^c) \\ \varphi_c(\tau_{1,1}^c)\varphi_c(\tau - y) d\tau_{1,1}^c dy d\tau \end{array} \right) \\
&= \int_0^t S_d(t - \tau) S_u(t - \tau) \int_0^\tau \varphi_u(a) S_d(a) S_x(\tau - a) \int_0^{\tau - a} \varphi_c(\tau_{1,1}^c)\varphi_c\left(\tau - a - \tau_{1,1}^c\right) d\tau_{1,1}^c da d\tau \\
&= \int_0^t S_d(t - \tau) S_u(t - \tau) \int_0^\tau \varphi_u(a) S_d(a) \varphi_{2,c}(\tau - a) S_x(\tau - a) da d\tau
\end{aligned}
\tag{A-4}
$$

The fifth equality of Eq. (A-4) follows from the change of variables, $\tau_0^u + \tau_{1,1}^c = y$ and $\tau_0^u + \tau_{1,1}^c + \tau_{1,2}^c = w$. The ninth equality of Eq. (A-4) is obtained by the change of variable $y - \tau_{1,1}^c = a$. The last equality of Eq. (A-4) is defined as $\varphi_{2,c}(t) = \int_0^t \varphi_c(\tau)\varphi_c(t-\tau)d\tau$. We define $\varphi_{c,x}(t)$, $\varphi_{u,d}(t)$, and $S_{d,u}(t)$ as $\varphi_{c,x}(t) \equiv \varphi_{j,c}(t)S_x(t)$ via $\varphi_{j,c}(t)$, that is, the probability density function of the time required for viral genetic material to produce *j* virions, or the *j*-fold convolution of $\varphi_c(t)$, $\varphi_{u,d}(t) \equiv \varphi_u(t)S_d(t)$, and $S_{d,u}(t) \equiv S_d(t)S_u(t)$. From repeating a similar procedure to derive Eqs. (A-3) and (A-4), we obtain the mean of the second term on the RHS of Eq. (A-1):

$$\theta(\tau_0^d - \tau_0^u)\sum_{j_1=1}^{\infty}\theta(t - t_{1,j_1}^c)\theta(t_{1,j_1}^d - t)\theta(t_{1,j_1}^u - t)\theta\left(\tau_0^x - \sum_{l=1}^{j_1}\tau_{1,l}^c\right) = L^{-1}\left(\hat{\varphi}_{u,d}(s)\left(\sum_{j_1=1}^{\infty}\hat{\varphi}_{j_1,c,x}(s)\right)\hat{S}_{d,u}(s)\right), \quad \text{(A-5)}$$

where $\hat{\varphi}(s)$ denotes the Laplace transform of $\varphi(t)$ ($\hat{\varphi}(s) = \int_0^{\infty}\varphi(t)e^{-st}$), $L^{-1}$ represents the inverse Laplace transform, and $\hat{\varphi}_{j,c,x}(s)$ indicates the Laplace transform of $\varphi_{j,c,x}(t) \equiv \varphi_{j,c}(t)S_x(t)$. The average of the final term on the RHS of Eq. (A-1) can be derived using a method similar to the derivation of Eq. (A-5), as follows:

$$\begin{aligned}
&\theta(\tau_0^d - \tau_0^u)\sum_{i=2}^{\infty}\sum_{j_1=1}^{\infty}\sum_{j_2=1}^{\infty}\cdots\sum_{j_i=1}^{\infty}\left\{\begin{aligned}&\left(\prod_{m=1}^{i-1}\theta(\tau_{m,j_m}^d - \tau_{m,j_m}^u)\right)\theta(t - t_{i,j_i}^c)\theta(t_{i,j_i}^u - t)\theta(t_{i,j_i}^d - t)\\ &\qquad\left(\prod_{q=0}^{i-2}\theta\left(\tau_{q,j_q}^x - \sum_{r=1}^{b}\tau_{q+1,y_r}^c\right)\right)\theta\left(\tau_{\mathrm{i}-1,j_{i-1}}^x - \sum_{l=1}^{g}\tau_{q+1,z_l}^c\right)\end{aligned}\right\}, \quad \text{(A-6)}\\
&= L^{-1}\left[\hat{S}_{d,u}(s)\sum_{i=2}^{\infty}\left\{\hat{\varphi}_{u,d}^i(s)\left(\sum_{j=1}^{\infty}\hat{\varphi}_{j,c,x}(s)\right)^i\right\}\right]
\end{aligned}$$

The sum of Eqs. (A-2), (A-5), and (A-6) completes the proof of Eq. (1) in the main text.